\documentclass{emulateapj}

\usepackage{amssymb}
\usepackage{bm}

\begin{document}

\title{Kinetic Alfv\'{e}n turbulence below and above ion-cyclotron frequency}
\author{J. S. Zhao$^{1,2}$, Y. M.~Voitenko$^{3}$, D. J. Wu$^{1}$ and
M.~Y.~Yu$^{4,5}$}

\begin{abstract}
Alfv\'{e}nic turbulent cascade perpendicular and parallel to the background
magnetic field is studied accounting for anisotropic dispersive effects and
turbulent intermittency. The perpendicular dispersion and intermittency make
the perpendicular-wavenumber magnetic spectra steeper and speed up
production of high ion-cyclotron frequencies by the turbulent cascade. On
the contrary, the parallel dispersion makes the spectra flatter and
decelerate the frequency cascade above the ion-cyclotron frequency.
Competition of the above factors results in spectral indices distributed in
the interval [-2,-3], where -2 is the index of high-frequency space-filling
turbulence, and -3 is the index of low-frequency intermittent turbulence
formed by tube-like fluctuations. Spectra of fully intermittent turbulence
fill a narrower range of spectral indices [-7/3,-3], which almost coincides
with the range of indexes measured in the solar wind. This suggests that the
kinetic-scale turbulent spectra are shaped mainly by dispersion and
intermittency. A small mismatch with measured indexes of about 0.1 can be
associated with damping effects not studied here.
\end{abstract}


\affil{1 Purple Mountain Observatory, Chinese Academy of Sciences,
Nanjing 210008, China. js\_zhao@pmo.ac.cn}

\affil{2 Key Laboratory of Solar activity, National Astronomical Observatories, Chinese
Academy of Sciences, Beijing 100012, China.}

\affil{3 Solar-Terrestrial Centre of Excellence, Space Physics Division,
Belgian Institute for Space Aeronomy, Ringlaan-3-Avenue Circulaire, B-1180
Brussels, Belgium}

\affil{4 Institute for Fusion Theory and Simulation and Department of Physics, Zhejiang
University, Hangzhou 310027, China.}
\affil{5 Institute for Theoretical Physics I, Ruhr University,
D-44780 Bochum, Germany}



\section{Introduction}

Alfv\'{e}nic turbulence, measured in situ by satellites in the Earth's space
environment (Chaston et al. 2008, 2009; Huang et al. 2012, 2014) and in the
solar wind (Sahraoui et al. 2010; He et al. 2011, 2013; Alexandrova et al.
2012; Horbury et al. 2012; Salem et al. 2012; Podesta 2013; Roberts et al.
2013, 2015), extends from magnetohydrodynamic (MHD) scales down to ion and
even electron kinetic scales. Turbulence at kinetic scales is often referred
to as kinetic Alfv\'{e}n turbulence. Theoretical models have shown that
kinetic Alfv\'{e}n turbulence is generated naturally through an anisotropic
Alfv\'{e}n wave cascade produced by interactions among counterstreaming Alfv%
\'{e}n wave packets (e.g., Kraichnan 1965; Goldreich \& Sridhar 1995;
Quataert \& Gruzinov 1999; Cranmer \& van Ballegooijen 2003; Schekochihin et
al. 2009; Bian et al. 2010; Howes et al. 2011b; Zhao et al. 2013).
Transition of Alfv\'{e}nic turbulence from MHD to kinetic scales depends
strongly on the plasma thermal to magnetic pressure ratio ($\beta $). For
example, the low-frequency turbulent cascade first arrives to the ion
gyroradius scale in $\beta >Q$ plasmas and to the electron inertial scale in
$\beta <Q$ plasmas, where $Q=m_{e}/m_{i}$ is the electron to ion mass ratio
(e.g., Zhao et al. 2013).

When the cascade reaches kinetic scales, the Kolomogorov-like spectrum $%
k_{\perp }^{-5/3}$ of perpendicular magnetic and electric fluctuations
transforms into the steeper $k_{\perp }^{-7/3}$ for the magnetic spectrum
and the flatter $k_{\perp }^{-1/3}$ for the electric spectrum (e.g.,
Schekochihin et al. 2009). The anisotropic scaling $k_{z}\propto k_{\perp
}^{2/3}$ at MHD scales becomes $k_{z}\propto k_{\perp }^{1/3}$ (for $\beta >Q
$) and $k_{z}\propto k_{\perp }^{7/3}$ (for $\beta <Q$) at kinetic scales
(Zhao et al. 2013), where $k_{z}\ $and $k_{\perp }$ are the wavenumbers
parallel and perpendicular to the background magnetic field $\mathbf{B}_{0}$%
. Therefore, with the increasing $k_{\perp }$ the parallel wavenumber $k_{z}$
and the Alfv\'{e}n wave frequency $\omega \ \left( \sim V_{A}k_{z}\right) $
are also increasing, where $V_{A}$ is the Alfv\'{e}n speed. When the
parallel wave scale approaches the ion inertial length $\lambda _{i}$, the
Alfv\'{e}n wave frequency approaches the ion-cyclotron frequency $\omega
_{ci}$, $\omega /\omega _{ci}\sim \lambda _{i}k_{z}\sim 1$. At this point
the low-frequency $\omega <\omega _{ci}$ kinetic Alfv\'{e}n turbulence
transforms into the high-frequency $\omega \gtrsim \omega _{ci}$ kinetic Alfv%
\'{e}n (sometimes named ``quasi-perpendicular whistler") turbulence.
Observations support the idea that high-frequency Alfv\'{e}n waves can be
generated by low-frequency Alfv\'{e}n waves via turbulent cascade (Huang et
al. 2012). Linear theories also agree that the kinetic Alfv\'{e}n branch can
extend from low- to high-frequency domain (Sahraoui et al. 2012; V\'{a}%
sconez et al. 2014; Zhao et al. 2014b).

It is not yet certain what effects make observed turbulent spectra $\propto
k_{\perp }^{-2.8}$\ (Alexandrova et al. 2012; Sahraoui et al. 2013) steeper
than the regular turbulent spectrum of KAWs $\propto k_{\perp }^{-7/3}$\
(see Schekochihin et al. 2009, and references therein). Two feasible
mechanisms discussed recently are Landau damping (Howes et al. 2011) and
intermittency (Boldyrev \& Perez 2012). If the turbulent cascade generates
high-frequency KAW, the spectra can be modified by dispersive effects of
finite $\lambda _{i}k_{z}$. High-frequency KAWs undergo also ion-cyclotron
wave-particle interactions (e.g., Voitenko \& Goossens 2002), which can
contribute to the turbulent spectra in addition to Landau damping. Kinetic
damping leads to the plasma heating and particles acceleration both along
and across $B_{0}\parallel z$, connecting particles to waves in the
solar-terrestrial environments (Marsch 2006). High-frequency effects in Alfv%
\'{e}nic turbulence are understood much less than the low-frequency ones.

Here we use a two-fluid plasma model to investigate dispersive effects (in
particular, of finite $\omega /\omega _{ci}\sim $\ $\lambda _{i}k_{z}$) and
intermittency in Alfv\'{e}nic turbulence and explore the wavenumber and
frequency spectra from MHD to kinetic scales. Influence of damping on the
turbulent spectra is not taken into account, which restricts applicability
of our results to the cases where spectral modifications due to damping are
weak in comparison to the modifications due to dispersion and intermittency.
In Discussion we argue this is often the case in the solar wind.

In the next Section we introduce a model for anisotropic Alfv\'{e}nic
turbulence and corresponding wavenumber spectra. The steady spectra in the
frequency space are presented in Section 3. Effects introduced by the
turbulent intermittency are described in Section 4. Section 5 presents the
spectral distributions in solar flare loops and in solar wind at 1 AU.
Section 6 discusses impacts of the injection scales, intermittency and
dissipative effects. The last section presents a summary of obtained
results. In Appendix A the analytical expressions of the wave variables and
their spectral distributions are derived for low-$\beta $ plasmas, and in
Appendix B the analytical results are extended to the case of intermittent
turbulence.

\section{Anisotropic Alfv\'{e}nic turbulence}

The steady-state spectral properties of Alfv\'{e}nic turbulence can be
investigated phenomenologically using a model developed by (e.g.,
Schekochihin et al. 2009), where turbulence is stirred initially through
collisions of counterstreaming Alfv\'{e}n wave-packets at the (spatial and
temporal) MHD scales. In the collisions, the local wave-wave interaction
condition as well as the critical balance condition are assumed to be
satisfied. The latter corresponds to strong Alfv\'{e}nic turbulence where
the nonlinear timescale becomes as short as the linear timescale.

To obtain the steady-state spectrum, one makes use of linear responses of
the Alfv\'{e}n mode, namely the quasilinear premise (Schekochihin et al.
2009; Howes et al. 2011a; Zhao et al. 2013). Therefore, for the spectrum of
the high-frequency kinetic-scale Alfv\'{e}nic turbulence, one needs the
linear relations for high-frequency kinetic Alfv\'{e}n wave (KAW). Recently,
Zhao et al. (2014b) derived the linear KAW responses at parallel
length-scales extending down to the ion inertial length and below, with
corresponding frequencies extending to $\omega \sim \omega _{ci}$ and above.
Below we use these results to obtain the high-frequency Alfv\'{e}nic
turbulent spectra and scalings.

In the case of local nonlinear interactions, the spectral energy flux can be
written as \citep{zhao2013a},
\begin{equation}
\epsilon =C_{1}^{-3/2}k_{\perp }\delta v_{e\perp }\delta b_{\perp }^{2},
\label{1}
\end{equation}%
where $C_{1}$ is of order unity, $\delta v_{e\perp }$ is the perpendicular
electron velocity fluctuation, and $\delta b_{\perp }$ is the perpendicular
magnetic fluctuation in velocity units, $\delta b_{\perp }\equiv V_{A}\delta
B_{\perp }/B_{0}$. Using the relation
\[
\delta v_{e\perp }=\bar{\omega}\mathcal{L}\delta b_{\perp },
\]%
we get
\begin{equation}
\delta b_{\perp }=C_{1}^{1/2}\epsilon ^{1/3}\bar{\omega}^{-1/3}k_{\perp
}^{-1/3}\mathcal{L}^{-1/3},  \label{2}
\end{equation}%
where the normalized frequency of oblique Alfv\'{e}n waves is (Zhao et al.
2014b)
\begin{equation}
\bar{\omega}\equiv \left[ \frac{\mathcal{R}+2\beta }{2\left( \mathcal{L}%
^{\prime }+\mathcal{L}^{2}\beta \right) }\left( 1+\sqrt{1-4\beta \frac{%
\mathcal{L}^{\prime }+\mathcal{L}^{2}\beta }{\left( \mathcal{R}+2\beta
\right) ^{2}}}\right) \right] ^{1/2},  \label{3}
\end{equation}%
with the definitions $\beta =k_{B}\left( T_{i}+T_{e}\right) /\left(
m_{i}V_{A}^{2}\right) $, $\mathcal{R}\equiv 1+\rho ^{2}k_{\perp }^{2}$, $%
\mathcal{L}\equiv 1+\lambda _{e}^{2}k_{\perp }^{2}$, $\mathcal{L}^{\prime
}\equiv 1+\lambda _{e}^{2}k_{\perp }^{2}+\lambda _{i}^{2}k_{z}^{2}$, $\rho =%
\sqrt{k_{B}\left( T_{i}+T_{e}\right) /m_{i}}/\omega _{ci}$, $\lambda
_{i}=V_{A}/\omega _{ci}$, and $\lambda _{e}=\sqrt{m_{e}/m_{i}}\lambda _{i}$.
The Alfv\'{e}n wave dispersion $\omega =V_{A}k_{z}\bar{\omega}$ is valid at
the quasi-perpendicular propagations, $k_{\perp }^{2}/k_{z}^{2}\gg 1$, in
the MHD and kinetic ranges without restrictions on $\omega /\omega _{ci}$.

From the critical balance condition that the linear Alfv\'{e}n time is equal
to the nonlinear turnover time \cite{gs1995},
\[
\omega ^{-1}=\left( C_{2}\delta v_{e\perp }k_{\perp }\right) ^{-1},
\]%
we present the anisotropy scaling relation as
\begin{equation}
k_{z}=C_{1}^{1/2}C_{2}V_{A}^{-1}\epsilon ^{1/3}\bar{\omega}^{-1/3}k_{\perp
}^{2/3}\mathcal{L}^{2/3},  \label{4}
\end{equation}%
where $C_{2}$ is a constant of order unity. Note since the magnetic field is
frozen into the electron before the turbulence cascading into electron
gyroradius scale, the electron velocity $\delta v_{e\perp }$ is used to
estimate the nonlinear turnover time.

Linear relations between electric and magnetic fluctuations are given by
\[
\delta e_{\perp }=\bar{\omega}\Gamma \delta b_{\perp }
\]%
and
\[
\delta e_{z}=\bar{\omega}\left( \Gamma -1\right) \left( k_{z}/k_{\perp
}\right) \delta b_{\perp },
\]%
where $\delta e_{\perp }\equiv \delta E_{\perp }/B_{0}$ and $\delta
e_{z}\equiv \delta E_{z}/B_{0}$. Together with Equations (\ref{2}) and (\ref%
{4}) one can then obtain the scaling relations for the electric components:
\begin{equation}
\delta e_{\perp }=C_{1}^{1/2}\epsilon ^{1/3}k_{\perp }^{-1/3}\bar{\omega}%
^{2/3}\mathcal{L}^{-1/3}\Gamma ,  \label{5}
\end{equation}%
and
\begin{equation}
\delta e_{z}=C_{1}C_{2}V_{A}^{-1}\epsilon ^{2/3}k_{\perp }^{-2/3}\bar{\omega}%
^{-1/3}\mathcal{L}^{1/3}\left( \Gamma -1\right) ,  \label{6}
\end{equation}%
where $\Gamma \equiv \mathcal{L}\widetilde{T}_{i}+\left( 1/\bar{\omega}%
^{2}-\lambda _{i}^{2}k_{z}^{2}\right) \widetilde{T}_{e}$ and $\widetilde{T}%
_{e,i}\equiv T_{e,i}/\left( T_{e}+T_{i}\right) $. The parallel electric
field $\delta e_{z}$ in Equation (6) is usually much smaller than the
perpendicular electric field $\delta e_{\perp }$ in Equation (5), but $%
\delta e_{z}$ can play an important role in the particles energization along
$\mathbf{B}_{0}$.

From Equations (2) and (5) we obtain the following magnetic and electric
power spectra:
\begin{eqnarray}
P_{\delta b_{\perp }} &=&k_{\perp }^{-1}\delta b_{\perp }^{2}  \nonumber \\
&=&C_{1}\epsilon ^{2/3}\bar{\omega}^{-1/3}k_{\perp }^{-5/3}\mathcal{L}%
^{-1/3},  \label{7}
\end{eqnarray}%
\begin{eqnarray}
P_{\delta e_{\perp }} &=&k_{\perp }^{-1}\delta e_{\perp }^{2}  \nonumber \\
&=&C_{1}\epsilon ^{2/3}\bar{\omega}^{4/3}k_{\perp }^{-5/3}\mathcal{L}%
^{-2/3}\Gamma ^{2}.  \label{8}
\end{eqnarray}

The physical quantities and spectra (\ref{2})--(\ref{8}) reduce to their MHD
counterparts when the kinetic factors $\rho k_{\perp }$, $\lambda
_{e}k_{\perp }$ and $\lambda _{i}k_{z}$ vanish.\ In low-$\beta $ $\left(
\beta \ll 1\right) $ plasmas, expressions (\ref{2})--(\ref{8}) can be
simplified (see in Appendix A).\ Note that the turbulence scalings in
plasmas $\beta \sim 1$ is nearly the same as that in $Q\ll \beta \ll 1$ due
to the similar properties of KAWs in these plasma environments.

Properties of the spectral scalings and physical quantities are summarized
in Table 1 for three $\beta $ regimes: inertial ($\beta \ll Q$), transition (%
$\beta \sim Q$), and kinetic ($Q\ll \beta <1$). Electron and ion
temperatures are assumed to be equal, $T_{i}=T_{e}$, so that $\rho \simeq
1.4\rho _{i}\simeq 1.4\rho _{s}$, where $\rho _{i}=\sqrt{k_{B}T_{i}/m_{i}}%
/\omega _{ci}$\ is the ion gyroradius, and $\rho _{s}=\sqrt{k_{B}T_{e}/m_{i}}%
/\omega _{ci}$\ is the ion-acoustic gyroradius.

At MHD scales, where all $\rho k_{\perp },\lambda _{e}k_{\perp }$, and $%
\lambda _{i}k_{z}$ are small, the scalings shown in Table 1 are consistent
with those described by \cite{gs1995} for the strong MHD Alfv\'{e}nic
turbulence. As the turbulence cascades into the kinetic scales, there can be
two cases: (i) for the large perpendicular kinetic effect $\lambda
_{e}k_{\perp }>\lambda _{i}k_{z}$ in the inertial and transition regimes,
and $\rho k_{\perp }>1$ and $\lambda _{i}k_{z}\lesssim 1$ in the kinetic
regime (outside the parenthesis), and (ii) for the large parallel kinetic
effect $\lambda _{i}k_{z}>\lambda _{e}k_{\perp }$ in the inertial and
transition regimes, and $\lambda _{i}k_{z}>\rho k_{\perp }$ in the kinetic
regime (inside the parenthesis). In the latter case, the wave frequency is
larger than the ion cyclotron frequency, and two limits $\lambda
_{e}k_{\perp }<\lambda _{i}k_{z}\ll \left( Q/\beta \right) ^{1/2}$\ and $%
\lambda _{i}k_{z}\gg \max \left( \lambda _{e}k_{\perp },\left( Q/\beta
\right) ^{1/2}\right) $\ are used to consider two different effects in
contributing the parallel electric field $\delta e_{z}$ in Eq. (A3). We see
that $k_{z}\propto k_{\perp }^{7/3}\ (k_{\perp }^{3})$ in the inertial
regime where $\lambda _{e}k_{\perp }\gg 1\gg \rho k_{\perp }$, and $%
k_{z}\propto k_{\perp }^{2}\ (k_{\perp }^{5/2})$ in the transition regime
where $\lambda _{e}k_{\perp }\sim \rho k_{\perp }\gg 1$, which means the
turbulence cascade proceeds mainly towards the $\mathbf{B}_{0}$ direction.
We found several new scalings, i.e., $P_{\delta b_{\perp }}\propto k_{\perp
}^{-1}$ in the inertial regime and $P_{\delta b_{\perp }}\propto k_{\perp
}^{-2}$ in the kinetic regime. We also note that the scalings in the
inertial regime at $\rho k_{\perp }\gg 1$, as well as in the kinetic regime
at $\lambda _{e}k_{\perp }\gg 1$, are the same as that in the transition
regime.

\begin{table*}[t]
\caption{Turbulence scalings. Scalings above the ion-cyclotron frequency are
shown in parenthesis. }
\begin{ruledtabular}
\begin{tabular}{ccccccc}
& \multicolumn{2}{c}{Inertial regime $(\beta \ll Q)$}
& \multicolumn{2}{c}{Transition regime $(\beta \sim Q)$}
& \multicolumn{2}{c}{Kinetic regime $(Q\ll \beta \ll 1)$}
\\
Parameter
& $1/k_{\perp }\gg \lambda _{e}$
& $\lambda _{e}\gg 1/k_{\perp}\gg \rho $
& $1/k_{\perp }\gg \lambda _{e}\sim \rho $
& $\lambda _{e}\sim\rho \gg 1/k_{\perp }$
& $1/k_{\perp }\gg \rho $
& $\rho \gg 1/k_{\perp }\gg\lambda _{e}$
\\ \hline
$P_{\delta b_{\perp }}(k_{\perp })$
& $k_{\perp }^{-5/3}$
& $k_{\perp}^{-7/3}(k_{\perp }^{-1})$
& $k_{\perp }^{-5/3}$
& $k_{\perp }^{-3}(k_{\perp}^{-2})$
& $k_{\perp }^{-5/3}$
& $k_{\perp }^{-7/3}(k_{\perp }^{-2})$
\\
$P_{\delta e_{\perp }}(k_{\perp })$
& $k_{\perp }^{-5/3}$
& $k_{\perp}^{-1/3}({k_{\perp }^{-3}}\tablenotemark{a}, {k_\perp^9}\tablenotemark{b})$
& $k_{\perp }^{-5/3}$
& $k_{\perp }(k_{\perp}^{5})$
& $k_{\perp }^{-5/3}$
& $k_{\perp }^{-1/3}(k_{\perp })$
\\
$\delta e_{z}(k_{\perp })$
& $k_{\perp }^{4/3}$
& $k_{\perp }^{5/3}({k_{\perp }}\tablenotemark{a},{k_{\perp }^{7}}\tablenotemark{b})$
& $k_{\perp}^{4/3}$
& $k_{\perp }^{2}(k_{\perp }^{9/2})$
& $k_{\perp }^{4/3}$
& $k_{\perp }^{-1/3}(k_{\perp }^{1/2})$
\\
$k_{z}(k_{\perp })$
& $k_{\perp }^{2/3}$
& $k_{\perp }^{7/3}(k_{\perp }^{3})$
& $k_{\perp }^{2/3}$
& $k_{\perp }^{2}(k_{\perp }^{5/2})$
& $k_{\perp}^{2/3} $
& $k_{\perp }^{1/3}(k_{\perp }^{1/2})$
\\
&  &  &  &  &  &  \\
$\omega (k_{\perp })$
& $k_{\perp }^{2/3}$
& $k_{\perp }^{4/3}(\omega
\rightarrow \omega _{ci})$
& $k_{\perp }^{2/3}$
& $k_{\perp }^{2}(k_{\perp}) $
& $k_{\perp }^{2/3}$
& $k_{\perp }^{4/3}(k_{\perp })$
\\
$P_{\delta b_{\perp }}(\omega )$
& $\omega ^{-2}$
& $\omega ^{-2}$
& $\omega^{-2}$
& $\omega ^{-2}(\omega ^{-2})$
& $\omega ^{-2}$
& $\omega^{-2}(\omega ^{-2})$
  \\
$P_{\delta e_{\perp }}(\omega )$
& $\omega ^{-2}$
& $\omega ^{-1/2}$
& $\omega ^{-2}$
& $\omega ^{0}(\omega ^{5})$
& $\omega ^{-2}$
& $\omega^{-1/2}(\omega )$
\end{tabular}
\footnotetext{~a~for $\lambda_ek_\perp<\lambda_ik_z \ll (Q/\beta)^{1/2}$.}
\footnotetext{~b~for
$\lambda_ik_z\gg\mathrm{max}(\lambda_ek_\perp,(Q/\beta)^{1/2}$).}
\end{ruledtabular}
\end{table*}

\section{Frequency spectra}

Accounting for the anisotropy relation (4), we can write the scaling of the
wave frequency as
\begin{equation}
\omega =C_{1}^{1/2}C_{2}\epsilon ^{1/3}\bar{\omega}^{2/3}k_{\perp }^{2/3}%
\mathcal{L}^{2/3},  \label{9}
\end{equation}%
whose asymptotic forms in low-$\beta $ plasmas are given in Table 1. The
power spectra $P_{\delta b_{\perp }}\left( \omega \right) $\ and\ $P_{\delta
e_{\perp }}\left( \omega \right) $ are found from the energy balance
condition $k_{\perp }P_{\delta b_{\perp },e_{\perp }}\left( k_{\perp
}\right) =\omega P_{\delta b_{\perp },e_{\perp }}\left( \omega \right) $,
which also presented in Table 1. Note that in the inertial regime the wave
frequency approaches $\omega _{ci}$ as $\lambda _{i}k_{z}>\lambda
_{e}k_{\perp }>1$. It shows that the universal spectrum $P_{\delta b_{\perp
}}(\omega )\propto \omega ^{-2}$ arises in all ranges, whereas $P_{\delta
e_{\perp }}(\omega )$ varies in different limits.\textbf{\ }The universal
frequency spectrum of magnetic power is not a surprising result, it is
rather a direct consequence of the constant energy flux and critical balance
that we assume here.

\section{Impact of intermittency on turbulent spectra and production of high
frequencies}

Our previous analysis assumed a space-filling turbulence where the turbulent
fluctuations of any particular scale cover all the volume occupied by the
turbulence. However, Alfv\'{e}nic turbulence observed in solar-terrestrial
plasmas often exhibits a non space-filling, i.e. intermittent character
(e.g., Chaston et al. 2008; Huang et al. 2012; Wu et al. 2013; Chen et al.
2014). Recent PIC simulations also suggest a non space-filling Alfv\'{e}nic
turbulence at kinetic scales (Wu et al. 2013). Two-fluid simulations by
Boldyrev \& Perez (2012) and gyrokinetic simulations by TenBarge \& Howes
(2013) explored the intermittent KAW turbulence extending from ion to
electron scale. The spectral index $\sim -8/3$\ obtained by Boldyrev \&
Perez (2012) is similar to the spectrum $-2.8$\ obtained by TenBarge \&
Howes (2013) (also see Howes et al. 2011). Both above studies suggest that
intermittency affects the spectral scalings in kinetic Alfv\'{e}nic
turbulence.

Here, we further assume that the Alfv\'{e}nic turbulence is mostly
space-filling at the MHD scales but becomes intermittent at kinetic scales.
Adopting the same approach as was used by Boldyrev \& Perez (2012), in
Appendix B we obtain analytical expressions and steady-state spectra for
different kinds of the intermittent Alfv\'{e}nic turbulence. The
corresponding turbulence scalings are presented in Table 2, where two kinds
of intermittent structures are considered, sheet-like $\left( \alpha
=1\right) $\ and tube-like $\left( \alpha =2\right) $. The normalized
wavenumbers are ordered as $\lambda _{e}k_{\perp }\gg 1\gg \rho k_{\perp }$\
in the very low-$\beta $\ plasmas ($\beta <Q$); $\lambda _{e}k_{\perp }\sim
\rho k_{\perp }\gg 1$\ for the transition $\beta \sim Q$; and $\rho k_{\perp
}\gg 1\gg \lambda _{e}k_{\perp }$\ for $\beta >Q$.

The perpendicular wavenumber spectrum ($\tilde{P}_{\delta b_{\perp }}$ in
Table 2)\ is steeper in the intermittent turbulence than in the
space-filling turbulence ($P_{\delta b_{\perp }}$ in Table 1), such that $%
\tilde{P}_{\delta b_{\perp }}=k_{\perp }^{-\alpha /3}P_{\delta b_{\perp }}$.
The same concerns also electric spectra, $\tilde{P}_{\delta e_{\perp
}}=k_{\perp }^{-\alpha /3}P_{\delta e_{\perp }}$. As $\alpha $\ varies
between 1 and 2 depending on the fractional content of the sheet- and
tube-like intermittent fluctuations, the magnetic spectral index spans the
range $R_{\mathrm{int}}=\left[ -7/3,-3\right] $\ in the wavenumber range $%
\rho _{i}k_{\perp }>1$\ and $\lambda _{i}k_{z}\lesssim 1$\ at $Q\ll \beta
\ll 1$.\ We believe that the spectra are the same also in $\beta \sim 1$\
plasmas where KAW properties are similar to that at $Q\ll \beta \ll 1$. Such
$\beta $\ values are typical for the solar wind at 1AU.

It is interesting to note that the parallel wavenumber $k_{z}$\ and
frequency $\omega $\ spectra of the intermittent turbulence (Table 2) are
also steeper than the corresponding spectra of the space-filling turbulence
(Table 1). It means that the parallel turbulence scale and wave frequency in
the intermittent turbulence approaches the ion inertial length $\lambda _{i}$%
\ and ion cyclotron frequency $\omega _{ci}$\ faster than in the
space-filling turbulence. In other words, turbulent intermittency
facilitates generation of high-frequency KAWs by the turbulent cascade.
Table 2 also shows that the magnetic frequency spectrum retains its original
form $\propto \omega ^{-2}$, while the electric frequency spectrum varies
depending on the $\beta /Q$\ ratio.

Recently, Sahraoui et al. (2013) have analyzed magnetic spectra selected
from 10 years of Cluster observations. This analysis has shown that the
spectral index\ at scales between ion and electron gyroradii is distributed
in the range $\left[ -2.5,-3.1\right] $\ with a peak at about $-2.8$. This
range is almost the same as the range $R_{\mathrm{int}}$\ predicted for the
intermittent turbulence, which suggests that the kinetic-scale solar wind
turbulence is intermittent. A slight (about $0.1$)\ down-shift of the
measured range as compared to $R_{\mathrm{int}}$ may be caused by the
dissipative effects that are not taken into account in the present study.

\begin{table*}[tbp]
\caption{Turbulence scalings for intermittent kinetic Alfv\'{e}n turbulence.
Scalings above the ion-cyclotron frequency are shown in parenthesis. }%
\begin{ruledtabular}
\begin{tabular}{ccccccc}
& \multicolumn{2}{c}{Inertial regime $(\beta \ll Q)$}
& \multicolumn{2}{c}{Transition regime $(\beta \sim Q)$}
& \multicolumn{2}{c}{Kinetic regime $(Q\ll \beta \ll 1)$}
\\
Parameter
& sheet-like & tube-like
& sheet-like & tube-like
& sheet-like & tube-like
\\ \hline
$\tilde{P}_{\delta b_{\perp }}(k_{\perp })$
& $k_{\perp }^{-8/3}(k_{\perp}^{-4/3})$
& $k_{\perp }^{-3}(k_{\perp }^{-5/3})$
& $k_{\perp}^{-10/3}(k_{\perp }^{-7/3})$
& $k_{\perp }^{-11/3}(k_{\perp }^{-8/3})$
& $k_{\perp }^{-8/3}(k_{\perp }^{-7/3})$
& $k_{\perp }^{-3}(k_{\perp }^{-8/3})$
\\
$\tilde{P}_{\delta e_{\perp }}(k_{\perp })$
& $k_{\perp }^{-2/3}({k_{\perp}^{-10/3}}^{a}, {k_{\perp}^{26/3}}^{b})$
& $k_{\perp }^{-1}({k_{\perp }^{-11/3}}^{a},{k_{\perp }^{25/3}}^{b})$
& $k_{\perp}^{2/3}(k_{\perp }^{14/3})$
& $k_{\perp }^{1/3}(k_{\perp }^{13/3})$
& $k_{\perp }^{-2/3}(k_{\perp }^{2/3})$
& $k_{\perp }^{-1}(k_{\perp }^{1/3})$
\\
$\delta \tilde{e}_{z}(k_{\perp })$
& $k_{\perp }^{7/3}({k_{\perp }^{5/3}}^{a},{k_{\perp }^{23/3}}^{b})$
& $k_{\perp }^{3}({k_{\perp }^{7/3}}^{a},{k_{\perp }^{25/3}}^{b})$
& $k_{\perp }^{8/3}(k_{\perp }^{31/6})$
& $k_{\perp }^{10/3}(k_{\perp }^{35/6})$
& $k_{\perp }^{1/3}(k_{\perp }^{7/6})$
& $k_{\perp }^{}(k_{\perp }^{11/6})$
\\
$\tilde{k}_{z}(k_{\perp })$
& $k_{\perp }^{8/3}(k_{\perp }^{10/3})$
& $k_{\perp }^{3}(k_{\perp }^{11/3})$
& $k_{\perp }^{7/3}(k_{\perp }^{17/6})$
& $k_{\perp }^{8/3}(k_{\perp }^{19/6})$
& $k_{\perp }^{2/3}(k_{\perp }^{5/6})$
& $k_{\perp }(k_{\perp }^{7/6})$
\\
&  &  &  &  &  &  \\
$\tilde{\omega} (k_{\perp })$
& $k_{\perp }^{5/3}(k_{\perp }^{1/3})$
& $k_{\perp }^{2}(k_{\perp }^{2/3})$
& $k_{\perp }^{7/3}(k_{\perp }^{4/3})$
& $k_{\perp }^{8/3}(k_{\perp }^{5/3})$
& $k_{\perp }^{5/3}(k_{\perp }^{4/3})$
& $k_{\perp }^{2}(k_{\perp }^{5/3})$
\\
$\tilde{P}_{\delta b_{\perp }}(\omega )$
& $\omega ^{-2}\left( \omega^{-2}\right) $
& $\omega ^{-2}\left( \omega ^{-2}\right) $
& $\omega^{-2}\left( \omega ^{-2}\right) $
& $\omega ^{-2}(\omega ^{-2})$
& $\omega^{-2}\left( \omega ^{-2}\right) $
& $\omega ^{-2}(\omega ^{-2})$
\\
$\tilde{P}_{\delta e_{\perp }}(\omega )$
& $\omega ^{-4/5}\left( {\omega^{-8}}^a, {\omega^{28}}^b \right) $
& $\omega ^{-1}\left( {\omega ^{-5}}^a, {\omega^{13}}^b \right) $
& $\omega^{-2/7}\left( \omega ^{13/4}\right) $
& $\omega ^{-1/2}(\omega ^{11/5})$
& $\omega ^{-4/5}\left( \omega ^{1/4}\right) $
& $\omega ^{-1}\left( \omega^{-1/5}\right) $
\\
\end{tabular}
\footnotetext{~a~for $\lambda_ek_\perp<\lambda_ik_z \ll (Q/\beta)^{1/2}$.}
\footnotetext{~b~for
$\lambda_ik_z\gg\mathrm{max}(\lambda_ek_\perp,(Q/\beta)^{1/2}$).}
\end{ruledtabular}
\end{table*}

\section{Two examples}

\subsection{Solar flare loops}

\begin{figure}[tbp]
\centerline{\includegraphics[width=7.5cm]{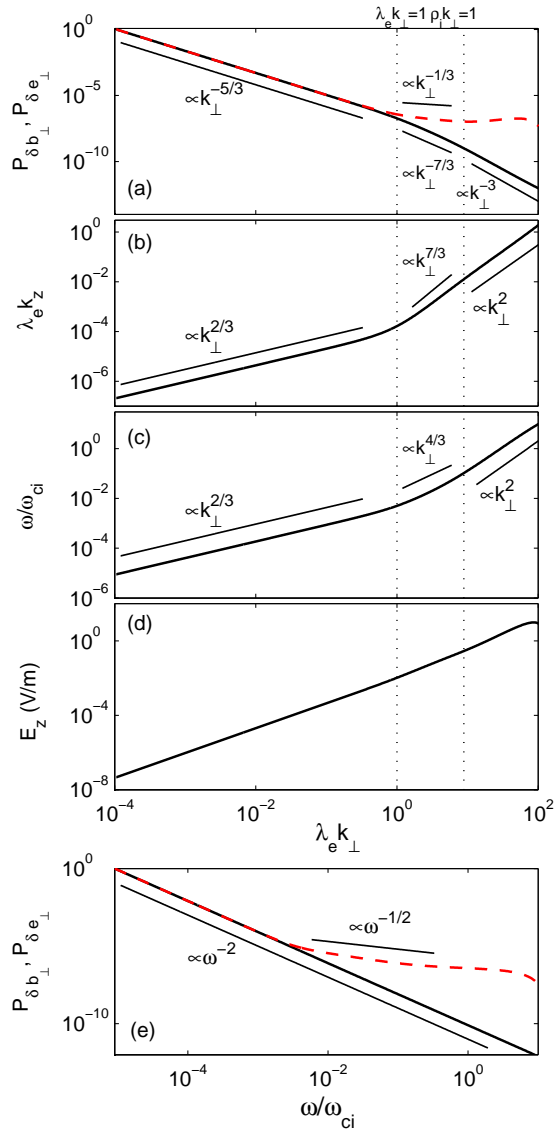}}
\caption{Steady-state spectral distributions in a typical flare loop plasma.
(a) Normalized energy spectra for $\protect\delta b_{\perp }$ (solid line),
and $\protect\delta e_{\perp }$ (dashed line). (b) The anisotropy relation.
(c) The wave frequency versus the perpendicular wavenumber. (d) The parallel
electric field versus the perpendicular wavenumber. (e) Normalized energy
spectra of $\protect\delta b_{\perp }$ (solid line) and $\protect\delta %
e_{\perp }$ (dashed line) versus the wave frequency.}
\label{fig:iner}
\end{figure}

Large nonthermal broadening of spectral lines, observed in the solar flares,
has been supposed to be produced by the plasma turbulence (e.g., Alexander
at al. 1998 and references therein). Let us consider excitation of the Alfv%
\'{e}nic turbulence in a solar flare loop with the length $L_{0z}=5\times
10^{6}$ m, width $L_{0\perp }=10^{4}$ m, internal magnetic field $%
B_{0}\simeq 5\times 10^{-2}$ T, number density $n_{0}\simeq 10^{15}$ m$^{-3}$%
, and temperature $T_{i}=T_{e}\simeq 10^{6}$ K \cite{zhao2013a}. We have
then $\beta \simeq 1.4\times 10^{-5}$, and the KAWs are in the inertial $%
\beta $ regime, $\beta <Q$. It is reasonable to assume that the initial
fluctuations perturbing the loop have the scales that are close to the loop
dimensions, $\lambda _{0\perp }\simeq L_{0\perp }$ and $\lambda _{0z}\simeq
L_{0z}$, such that $k_{0z}/k_{0\perp }=2\times 10^{-3}$.

The deduced from spectral observations non-thermal velocities $%
V_{nt}=100-200 $ km s$^{-1}$ (Alexander et al. 1998 and references therein)
allow estimating the turbulence amplitude $\delta B_{0\perp }$ at the
driving scale $k_{0\perp }$. Assuming that the turbulence is driven at MHD
scales, and using the MHD turbulent spectrum $\sim k_{\perp }^{-5/3}$, the
relation $V_{nt}^{2}\simeq V_{turb}^{2}$ gives
\begin{eqnarray}
\frac{V_{nt}^{2}}{V_{A}^{2}}=\frac{\delta B_{turb}^{2}}{B_{0}^{2}}=\frac{%
\delta B_{0\perp }^{2}}{B_{0}^{2}}\int_{k_{0\perp }}^{k_{d}}\frac{dk_{\perp }%
}{k_{0\perp }}\left( \frac{k_{\perp }}{k_{0\perp }}\right) ^{-5/3}  \nonumber
\\
=\frac{3}{2}\left( 1-\left( \frac{k_{d\perp }}{k_{0\perp }}\right)
^{-2/3}\right) \frac{\delta B_{0\perp }^{2}}{B_{0}^{2}},  \label{10}
\end{eqnarray}%
where the contribution originating from the dissipation scale $k_{d\perp }$
can be neglected for a wide inertial range $k_{d\perp }/k_{0\perp }\gg 1$.
Then, taking for certainty $V_{nt}=140$ km s$^{-1}$, we estimate $\delta
B_{0\perp }/B_{0}\simeq 2\times 10^{-3}$ and the perturbative approach is
justified. In this case, the critical balance condition $k_{0z}/k_{0\perp
}=\delta B_{0\perp }/B_{0}$ is satisfied already at the driving scales. The
initial wave frequency $\omega \simeq 43$ rad s$^{-1}$ is 5 orders below the
ion cyclotron frequency $\omega _{ci}\simeq 5\times 10^{6}$ rad s$^{-1}$.

Based on the above parameters, Figure 1 presents spectral scaling for the
turbulence in the flare loop. It is seen that the turbulent cascade arrives
to the high ion cyclotron frequencies at the perpendicular scales that are
already smaller than the ion gyroradius scale. Therefore, the scalings of
the kinetic Alfv\'{e}n turbulence are mainly defined by the perpendicular
dispersive effects of finite $\lambda _{e}k_{\perp }$, which result in the
magnetic spectrum $\propto k_{\perp }^{-7/3}$, electric spectrum $\propto
k_{\perp }^{-1/3}$, anisotropic scale relation $k_{z}\propto k_{\perp
}^{7/3} $, and frequency scaling $\omega \propto k_{\perp }^{4/3}$. It is
interesting to observe the increase of the parallel electric field $E_{z}$\
as the wavelength decreases, which attains large values $E_{z}\sim 0.1-1$\ $%
V/m$\ at kinetic scales. These values are larger than the Dreicer field $%
E_{D}=eln\Lambda /\left( 4\pi \epsilon _{0}\lambda _{D}^{2}\right) \sim
10^{-2}$ $V/m$, so that the turbulence-generated parallel electric fields
may play an important role in the field-aligned particle acceleration and/or
plasma heating in flare loops. One however should keep in mind that our
results are valid only in the wavenumber ranges where damping is relatively
weak (see Discussion). At high wavenumbers one may need to account for
dissipative effects (and for electron dispersive effects at $\rho
_{e}k_{\perp }\geq 1$).

The parallel dispersive effects are not expected to be significant in the
considered here case of the low-frequency driver perturbing the whole loop.
However, Alfv\'{e}nic perturbations excited by kinetic instabilities may be
high-frequency from the very beginning (see e.g. Voitenko and Goossens
2002), in which case the parallel dispersive effects of finite $\lambda
_{i}k_{z}$ must be taken into account. We also do not discuss here the
intermittency effects because it is difficult to deduce from the available
observations if the turbulence in flare loops is intermittent.

\subsection{Solar wind at 1 AU}

\begin{figure}[tpb]
\centerline{\includegraphics[width=7.5cm]{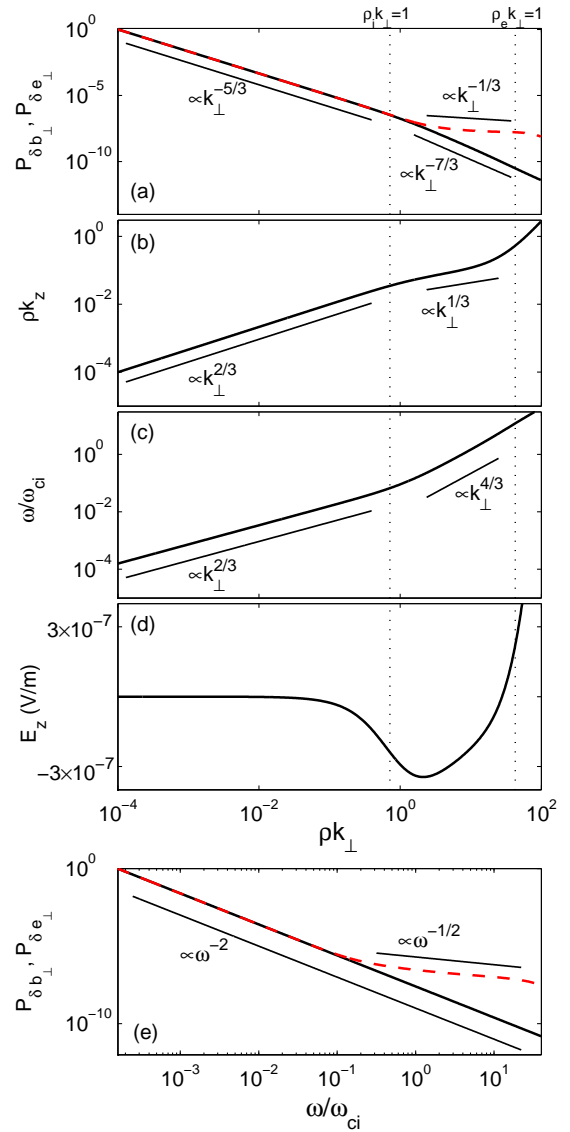}}
\caption{Steady-state Alfv\'en wave spectra in a typical solar wind plasma.
(a) Normalized energy spectra for $\protect\delta b_\perp$ (solid line), and
$\protect\delta e_\perp$ (dashed line). (b) The anisotropy relation. (c) The
wave frequency versus the perpendicular wavenumber. (d) The parallel
electric field versus the perpendicular wavenumber. (e) Normalized energy
spectra of $\protect\delta b_\perp$ (solid line) and $\protect\delta e_\perp$
(dashed line) versus the wave frequency. }
\label{fig:kine}
\end{figure}

Typical plasmas parameters are $B_{0}\simeq 11$ nT, $N_{0}\simeq 9\times
10^{6}$ m$^{-3}$, $T_{i}\simeq 1.5\times 10^{5}$ K and $T_{e}\simeq
1.4\times 10^{5}$ K in the solar wind at 1 AU \cite{sale2012}. Here $\beta
\sim 0.4$. We consider high-amplitude magnetic perturbations $\delta
B_{0\perp }\sim B_{0}$, at the isotropic initial (injection) scales $%
L_{0z}=L_{0\perp }=3\times 10^{9}$ m, which imply that the critical balance
is set up at the very beginning. The corresponding initial wave frequency $%
\omega \simeq 1.7\times 10^{-4}$ rad s$^{-1}$ is much less as compared to
the ion-cyclotron frequency $\omega _{ci}\simeq 1$ rad s$^{-1}$. Based on
above parameters, the spectral scalings are presented in Figure 2.

From Figure 2 we see that the wave frequency reaches the ion cyclotron
frequency well above the ion gyroradius scales. In this case kinetic effects
of finite $\rho _{i}k_{\perp }$ dominate kinetic spectral scalings and
result in the magnetic spectrum $\propto k_{\perp }^{-7/3}$, electric
spectrum $\propto k_{\perp }^{-1/3}$, anisotropy scale relation $%
k_{z}\propto k_{\perp }^{1/3}$, and frequency scaling $\omega \propto
k_{\perp }^{4/3}$ in the range $\rho _{i}k_{\perp }>1>\rho _{e}k_{\perp }$.
Although magnetic amplitudes are high at the injection scales, they drop
well below $B_{0}$ as the turbulence cascades towards the small scales, such
that $\delta B_{\perp }/B_{0}$ $\sim 0.05$ at the ion gyroradius scale $\rho
_{i}k_{\perp }=1$ and $\delta B_{\perp }/B_{0}$ $\sim 0.005$ at the electron
gyroradius scale $\rho _{e}k_{\perp }=1$. The sign \textquotedblleft $-$" in
front of $E_{z}$\ in Panel (d) represents the phase shift between $E_{z}$\
and $\delta B_{\perp }$.

However, the above scalings disagree with observed ones. Sahraoui et al.
(2013) have found that magnetic spectra $\propto k_{\perp }^{-7/3}$\ are
unlikely in the kinetic-scale solar wind turbulence. Most of the observed
kinetic-scale spectra are significantly steeper, with the spectral index
distributed in the range $\left[ -2.5,-3.1\right] $. The observed range of
spectral indices is very close to that predicted in the previous section for
the intermittent turbulence, $\left[ -7/3,-3\right] $\ (see also Table II).
This suggests that the kinetic-scale turbulence in the solar wind is
intermittent, with the steepest spectra dominated by the low-frequency
tube-like fluctuations, whereas the flattest spectra indicate the presence
of sheet-like structures and/or high frequencies. The intermittent character
of the solar wind turbulence at kinetic scales has been supported by recent
Cluster observations of magnetic (Wu et al. 2013) and density (Chen et al.
2014) fluctuations.

\section{Discussion}

\subsection{\textbf{Impact of the injection scales on the excitation of
high-frequency KAWs}}

\begin{figure}[tbp]
\centerline{\includegraphics[width=9.0cm]{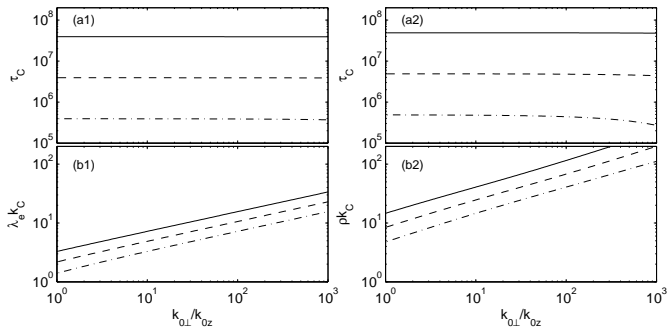}}
\caption{The spatial and temporal scales, $\protect\tau_c$ and $k_c$, for
different initial energy injection scale and anisotropy ratio. The solid,
dashed, and dash-dotted lines represent $\protect\lambda_ik_{0z}=10^{-5}$, $%
10^{-4}$ and $10^{-3}$, respectively. The plasma parameters used in panels
(a1) and (b1) are the same as that in figure 1; the plasma parameters in
panels (a2) and (b2) are the same as that in figure 2.}
\label{fig:iner}
\end{figure}

To further understand the frequency cascade in Alfv\'{e}nic turbulence, it
is of interest to estimate the time required for the wave frequency to
cascade from the initial driving frequency $\omega =\omega _{0}$ to the ion
cyclotron frequency $\omega =\omega _{ci}$. Assume that the perpendicular
wavenumber doubles at each ($j$-th) cascade step, $k_{\perp \left( j\right)
}=2k_{\perp \left( j-1\right) }$. The time of the $j$-th cascade step is $%
t_{j}=2\pi /\omega (k_{\perp j})$. If after $n$ cascades, at the critical
wavenumber $k_{\perp c}=k_{\perp \left( n\right) }$ the wave frequency
reaches $\omega _{ci}$, the time required is $\tau \sim \sum_{j=1}^{n}2\pi
/\omega (k_{\perp j})$. Figure 3 shows the normalized time scale $\tau
_{c}\equiv \tau /(2\pi /\omega _{ci})=\sum_{j=1}^{n}(\omega _{ci}/\omega
(k_{\perp j}))$ and the corresponding perpendicular wavenumber $k_{\perp c}$
as functions of the initial anisotropy $k_{0\perp }/k_{0z}$, for different
injection scales $\lambda _{i}k_{0z}$. We see that the wave frequency can
reach $\omega _{ci}$ in the vicinity of the plasma kinetic scales. Figure 3
also shows that larger $\lambda _{i}k_{0z}$ and $k_{0z}/k_{0\perp }$ leads
to shorter time $\tau _{c}$ and smaller wavenumber $k_{\perp c}$ at which
the ion cyclotron frequency is reached. Thus, detailed comparison of our
theoretical predictions with the satellite observations would require more
definite information on $\lambda _{i}k_{0z}$ and $k_{0\perp }/k_{0z}$, which
are not certain at present.

Earlier studies have shown that the ions of the solar atmosphere can be
accelerated by Alfv\'{e}n waves at the ion cyclotron frequency. It is
therefore of interest to consider the generation of such Alfv\'{e}n waves by
the turbulent cascade. We consider a plasma in the solar active region \cite%
{gary2001}, where $B_{0}=10^{-3}$ T, $n=10^{15}$ m$^{-3}$, and $T=3\times
10^{6}$ K, $\beta \simeq 0.1$ and $\omega _{ci}\simeq 10^{5}$ rad/s. The
times for the turbulent cascade to generate Alfv\'{e}n waves at the ion
cyclotron frequency are $\tau \sim 0.1$ s, 1 s, and 10 s for $\lambda
_{i}k_{0z}=10^{-3}$,$\ 10^{-4}$\ and $10^{-5}$, respectively. The
corresponding spatial scales are $L\sim V_{A}\tau \sim 10^{5}$ m, $10^{6}$
m, and $10^{7}$ m, respectively. These scales are smaller than the global
spatial scales $\sim 10^{8}-10^{9}$ m (temporal scales, ~ tens minutes) of
the active regions. Therefore, ion cyclotron Alfv\'{e}n waves may be easily
excited by the Alfv\'{e}nic turbulence cascade in the solar active regions,
where sources of the initial Alfv\'{e}n waves can be convective motions of
the magnetic foot-points, particle fluxes, and/or magnetic reconnection.

\subsection{\textbf{Impact of intermittency and spectra contamination by
high frequencies}}

Recent results on solar wind turbulence reported by Sahraoui et al. (2013)
indicate steep spectra at kinetic scales, with spectral indices distributed
in the interval $\left[ -2.5,-3.1\right] $. Such spectra can be formed by
the intermittent Alfv\'{e}nic turbulence, which is supported by the high
intermittency measured in the solar wind turbulence at kinetic scales (Wu et
al. 2013; Chen et al. 2014).\textbf{\ }Extending analysis below electron
gyroradius scale, Alexandrova et al. (2012) proposed a complex spectral form
$\propto k_{\perp }^{-8/3}\exp \left( -k_{\perp }\rho _{e}\right) $\ which
is nearly the power-law $\propto k_{\perp }^{-8/3}$\ between the ion and
electron gyroradius and mostly exponential below the electron gyroradius.
The power-low part of the spectrum ($\propto k_{\perp }^{-8/3}$)\ may be
formed by the intermittent turbulence with sheet-like structure (Boldyrev \&
Perez 2012), and steeper observed spectra can be formed by adding tube-like
structures. The exponential spectrum drop at $k_{\perp }\rho _{e}>1$\
implies the appearance of the dissipation range, and may be caused by the
strong electron Landau damping (TenBarge \& Howes 2013). On the other hand,
not exponential but steep power-law spectra below the electron gyroradius
scale were reported by Sahraoui et al. (2010, 2013), who also suggested
possible reasons for that. Further observations and analysis are needed to
distinguish the nature of turbulence at scales below $\rho _{e}$\ and to
resolve the mentioned above controversy. We did not study this range of
scales.

Direct comparisons between spacecraft-frame frequency spectra and
theoretical wavenumber spectra may be complicated by violation of the Taylor
hypothesis. Namely, as we have shown above, the KAW frequency can increase
to $\omega \gtrsim \omega _{ci}$, in which case the contribution of the term
$\sim \omega $\ to the spacecraft-frame frequency $\omega _{sc}=\omega
+k\cdot v_{sw}$\ may become as important as the contribution of Doppler term
$\sim k\cdot v_{sw}$. Therefore, to explain the broad index distribution
observed by Sahraoui et al. (2013) (see also Huang et al. 2014), one needs
further analysis of the possible production of high frequencies and their
contribution to the frequency spectra measured in the spacecraft frame.

It should also be noted that the anisotropy scaling is $k_{z}\propto
k_{\perp }^{2/3}$\ for intermittent low-frequency KAW turbulence formed by
the sheet-like fluctuations (Boldyrev \& Perez 2012).\

The mentioned above problems call for further investigations of the role of
intermittency in kinetic Alfv\'{e}n turbulence.

\subsection{\textbf{Damping effects}}

In the low-frequency kinetic Alfv\'{e}n turbulence, possessing parallel
electric fields, the proton and electron Landau damping may dissipate the
turbulent energy and influence spectral transfer. In this study we neglected
these dissipative effects, as well as the ion-cyclotron resonant damping,
which needs justifications. To this end we note that even the strongest
resonant damping, based on the Maxwellian velocity distributions, does not
prevent the super-ion-cyclotron KAWs from propagation. This can be directly
seen from Figs. 3-9 by Sahraoui et al. (2012) or Fig. 4 by V\'{a}sconez et
al. (2014) showing the KAW kinetic dispersion and damping. Even with fixed
Maxwellian velocity distributions, many dispersion curves, corresponding to
different propagation angles, extend continuously from sub- to
super-ion-cyclotron frequencies without being heavily damped. Say, the
relative damping rates in Fig. 9 by Sahraoui et al. (2012) are low, $\gamma
_{L}/\omega \lesssim 0.2$,\ for super-cyclotron KAWs at the propagation
angles $>80^{\circ }$. Hence the critically balanced turbulent cascade,
operating at the wave period time-scale $\gamma _{NL}\simeq \omega $, is
still much faster than the dissipation of super-cyclotron KAWs, $\gamma
_{NL}\gg \gamma _{L}$.\textbf{\ }

Influence of damping on the turbulent spectra can be strong. Cranmer \& van
Ballegooijen (2003) and Podesta et al. (2010) have shown that the magnetic
spectrum of low-frequency kinetic Alfv\'{e}n turbulence experiences a fast
fall-off between ion and electron scales if the Landau damping is accounted
for. The frequency spectrum $\propto \omega ^{-3.2}$, obtained in numerical
simulations by TenBarge \& Howes (2012) for $\beta =1$, is also much steeper
than our $\propto \omega ^{-2}$\ spectrum, which indicates a strong Landau
damping in their simulations. Indeed, Podesta et al. (2010) and TenBarge and
Howes (2012) used the linear Landau damping ($\gamma _{L}^{\mathrm{Maxwellian%
}}$)\ assuming Maxwellian velocity distributions of plasma species. However,
this approximation can hardly be applied to the solar wind where essentially
non-Maxwellian particle velocity distributions (PVDs) are regularly observed.

In accordance to recent analytical estimations (Voitenko \& De Keyser 2011;
Borovsky \& Gary 2011; Rudakov et al. 2011; Voitenko \& Pierrard 2013) and
numerical simulations (Pierrard \& Voitenko 2013; V\'{a}sconez et al. 2014),
the local velocity-space plateaus are formed in the solar wind PVDs by the
observed ion-scale turbulence. This conclusion is supported by many in-situ
observations of nonthermal features typical for such plateaus.\ Here we
refer to the recent paper by He et al. (2015) demonstrating clear
observational evidences of the quasilinear plateaus rendering $\gamma _{L}^{%
\mathrm{Maxwellian}}$ inappropriate. The real damping and its influence on
the observed spectra are therefore reduced by the particles feedback (see
equation (25) by Voitenko \& De Keyser 2011, and following discussions).

To obtain the average spectral index observed in the solar wind ($\sim -2.8$%
\ at $k_{\perp }\rho _{i}>1$) from the regular nondissipative spectral index
($\sim -7/3$),\ one needs to add the index decrement of about $-0.47$. In
accordance to recent findings, this decrement can be provided by damping
(Howes et al. 2011) and/or intermittency (Boldyrev et al. 2012). As the
sheet-like intermittency regularly appears in simulations (Boldyrev et al.
2012) and reduces the spectral index to $-8/3$, the rest $-0.13$\ can be
attributed to damping. The influence of damping on the turbulent spectra is
therefore small, in which case the two-fluid model is a good proxy to study
KAW turbulent spectra (see e.g. Boldyrev et al. 2012; V\'{a}sconez et al.
2014). Several damping mechanisms can contribute to the $-0.13$\ decrement,
including Landau damping, ion-cyclotron damping, non-adiabatic/stochastic
heating, etc. (see e.g. Quataert 1998; Voitenko \& Goossens 2004; Chandran
et al. 2010).

This conclusion is supported by the two-fluid simulations by Boldyrev \&
Perez (2012) giving the spectral index $\sim -8/3$, similar to the observed
indices and to the index $\sim -2.8$ found in gyrokinetic simulations (Howes
et al. 2011b; TenBarge \& Howes 2013). This suggests that the dissipative
effects are not so important for the turbulent spectra in the solar wind.

When the high-frequency $\omega \sim \omega _{ci}$ KAWs are excited by the
turbulent cascade, they can undergo the ion-cyclotron resonance with
particles satisfying the resonant condition $\omega -n\omega
_{ci}=k_{z}v_{z} $ (Hollweg \& Isenberg 2002; Voitenko \& Goossens 2002,
2003). However, this process depends not solely on the wave frequency, but
also on the wave polarization and plasma properties. In particular, both the
right-hand polarization of KAWs and the quasilinear modification of the
proton velocity distribution reduce the wave damping at the ion-cyclotron
resonance. The right-hand polarization allows also a smooth extension of the
KAW branch above the ion-cyclotron frequency (Boldyrev et al. 2013)\textbf{.
}In addition, the ion-cyclotron resonance is narrow-band (see e.g. Fig. 4 by
V\'{a}sconez et al. 2014), and the turbulent cascade can jump over the
narrow resonant layer and proceed further to higher frequencies. These
properties of KAWs explain their observations above the ion-cyclotron
frequency (Huang et al. 2012).

\subsection{\textbf{Roles of quasilinear premise and critical balance}}

To model the Alfv\'{e}nic turbulence we also used a quasilinear premise
(e.g., Schekochihin et al. 2009). Its validity has been supported by the
gyrokinetic simulations (i.e., Howes et al. 2011b). The quasilinear premise
has been widely used to distinguish the wave modes in the dissipation range
of the solar wind turbulence (e.g., Sahraoui et al. 2010; He et al. 2011;
Salem et al. 2012; Podesta 2013; Roberts et al. 2013, 2015), as well in
modeling the kinetic Alfv\'{e}n turbulence (i.e., Schekochihin et al. 2009;
Howes et al. 2011a; Voitenko and De Keyser 2011; Boldyrev \& Perez 2012;
Zhao et al. 2013). The related assumption of the critical balance between
the linear Alfv\'{e}n propagation timescale and the nonlinear turnover
timescales (Goldreich \& Sridhar 1995) is also regularly used in modeling
the strong Alfv\'{e}nic turbulence. Simulations of the kinetic Alfv\'{e}n
turbulence also support this assumption (e.g., Howes et al. 2011b).

Our model assumes local interactions among counter-propagating Alfv\'{e}nic
fluctuations forming the turbulence. Both these assumptions can be violated.
The nonlocal spectral transport (Voitenko \& Goossens 2005; Zhao et al.
2011a, 2011b, 2014a; Howes et al. 2011b) may contribute to the kinetic-scale
spectra and should be taken into account in more comprehensive models.
Furthermore, the turbulence generated by the nonlinear interaction between
co-propagating waves can produce much steeper spectra. As was shown by
Voitenko and De Keyser (2011), the nonlinear interactions among
co-propagating KAWs can produce steepest spectra in the vicinity of the ion
gyroscale, $P_{\delta b_{\perp }}\propto k_{\perp }^{-3}$ for the strong
turbulence and $P_{\delta b_{\perp }}\propto k_{\perp }^{-4}$ for the weak
turbulence. It is still unknown which interaction (the interaction between
counter-propagating KAWs or the interaction between co-propagationg KAWs) or
both domimates the kinetic Alfv\'{e}n turbulence observed in the solar wind.

At last, we note that the universal frequency spectrum of the magnetic power
$P_{\delta b_{\perp }}(\omega )\propto \omega ^{-2}$\ should not surprise
the reader. It is a rather natural consequence of the constant energy flux
and critical balance conditions and may change only if one or both of the
above conditions are violated. For example, if the wave damping becomes
strong at very small kinetic scales, the spectral energy flux deceases with $%
k_{\perp }$\ making spectra steeper.

\section{Summary}

We develop a semi-phenomenological model of Alfv\'{e}nic turbulence
extending from low frequencies $\omega \ll \omega _{ci}$ at MHD scales to
high frequencies $\omega \gtrsim \omega _{ci}$ at kinetic scales. The
quasi-stationary turbulent spectra are obtained accounting for the
dispersive effects of finite $\rho _{i}k_{\perp }$, $\lambda _{e}k_{\perp }$%
, and $\omega /\omega _{ci}\sim \lambda _{i}k_{z}$. New findings are
summarized as follows:

(1) Generation of high frequencies $\omega /\omega _{ci}\gtrsim 1$ by the
turbulent cascade is possible and depends on the injection scale, frequency,
and the turbulence state. Larger driving frequency $\omega _{0}/\omega _{ci}$
and anisotropy $k_{0z}/k_{0\perp }$ accelerate production of ion-cyclotron
frequencies $\omega \sim \omega _{ci}$. Large parallel wavenumbers and
frequencies are generated faster in the intermittent turbulence than in the
space-filling one, such that intermittent $\tilde{k}_{z}\propto k_{\perp
}^{\alpha /3}k_{z}$ and $\tilde{\omega}\propto k_{\perp }^{\alpha /3}\omega $%
, where $\alpha =1$ for the sheet-like intermittent structures and $\alpha
=2 $ for the tube-like.

(2) Parallel dispersive effects at $\lambda _{i}k_{z}\sim $ $\omega /\omega
_{ci}>1$ make kinetic-scale spectra and scalings flatter. In particular, the
magnetic spectral index is increased by 1/3 as compared to the low-frequency
KAW turbulence. In the space-filling turbulence, $P_{\delta b_{\perp
}}\propto k_{\perp }^{-2}$ for $Q\ll \beta \lesssim 1$, and $P_{\delta
b_{\perp }}\propto k_{\perp }^{-1}$ for $\beta \ll Q$.

(3) At $\lambda _{i}k_{z}\sim $ $\omega /\omega _{ci}\ \lesssim 1$, the
perpendicular dispersive effects of finite $\rho _{i}k_{\perp }$ (or $%
\lambda _{e}k_{\perp }$ at $\beta \ll Q$) dominate and the spectra remain
nearly the same as in the low-frequency kinetic Alfv\'{e}n turbulence, such
that $P_{\delta b_{\perp }}\propto k_{\perp }^{-7/3}$ and $P_{\delta
e_{\perp }}\propto k_{\perp }^{-1/3}$ in the space-filling turbulence.

(4) Magnetic and electric power spectra are steeper in the intermittent
turbulence ($\tilde{P}$) than in the space-filling ($P$): $\tilde{P}_{\delta
b_{\perp }}\propto k_{\perp }^{-\alpha /3}P_{\delta b_{\perp }}$ and $\tilde{%
P}_{\delta e_{\perp }}\propto k_{\perp }^{-\alpha /3}P_{\delta e_{\perp }}$
For the mixed sheet- and tube-like intermittency $1\leq \alpha \leq 2$,
which gives the range of possible spectral indices $R_{\mathrm{int}%
}=[-7/3,-3]$ for $Q\ll \beta \lesssim 1$.

(5) The universal frequency spectrum $E_{\delta b_{\perp }}\left( \omega
\right) \propto \omega ^{-2}$ is found. An apparent contradiction of this
universal spectrum with the non-universal effects of intermittency (see item
1 above) can be explained by the dominating cross-field dynamics governing
evolution of $k_{\perp }$, whereas frequency follows $k_{\perp }$ via
critical balance. In contrast to $E_{\delta b_{\perp }}\left( \omega \right)
$, the spectral index of electric frequency spectra $E_{\delta e_{\perp
}}\left( \omega \right) $ varies with the scale range, turbulence
intermittency, and plasma $\beta $.

(6) A good correspondence of $R_{\mathrm{int}}$ with the range of measured
spectral indexes $R_{\mathrm{obs}}=[-2.5,-3.1]$ suggest that the solar wind
turbulence at kinetic scale is intermittent and consists of varying
fractions of sheet-like and tube-like fluctuations. Shallower spectra may
indicate the presence of a fraction of high-frequency fluctuations.

Damping effects are not taken into account in our study, which is justified
if the dispersive and intermittency effects are stronger. In the solar wind
the turbulent spectral index $-2.8$\ is dominated by the inherent nonlinear
dynamics, and only small decrement of the index (about $-0.13$) can be
associated with damping. Weak damping effects can be consequence of the
quasilinear and/or nonlinear modifications of the particles velocity
distributions, which reduce damping at resonant scales $\rho _{i}k_{\perp
}>1 $.

\appendix

\section{Wave Variables and Spectra in Low-$\protect\beta $ Plasmas}

In low-$\beta $ ($\beta \ll 1$) plasmas, expressions (\ref{2}), (\ref{4})--(%
\ref{8}) can be simplified to
\begin{eqnarray}
\delta b_{\perp } &=&C_{1}^{1/2}\epsilon ^{1/3}k_{\perp }^{-1/3}\mathcal{R}%
^{-1/6}\mathcal{L}^{-1/3}\mathcal{L}^{\prime 1/6},  \label{A1} \\
\delta e_{\perp } &=&C_{1}^{1/2}\epsilon ^{1/3}k_{\perp }^{-1/3}\mathcal{R}%
^{-2/3}\mathcal{L}^{-1/3}\mathcal{L}^{\prime -1/3}\left[ \left( 1+\rho
_{i}^{2}k_{\perp }^{2}\right) \mathcal{L}-\rho _{s}^{2}k_{\perp }^{2}\lambda
_{i}^{2}k_{z}^{2}\right] ,  \label{A2} \\
\delta e_{z} &=&C_{1}C_{2}V_{A}^{-2}\epsilon ^{2/3}k_{\perp }^{4/3}\mathcal{R%
}^{-5/6}\mathcal{L}^{1/3}\mathcal{L}^{\prime -1/6}\left[ \rho _{s}^{2}\left(
1+\lambda _{i}^{2}k_{z}^{2}\right) -\lambda _{e}^{2}\left( 1+\rho
_{i}^{2}k_{\perp }^{2}\right) \right] ,  \label{A3} \\
k_{z} &=&C_{1}^{1/2}C_{2}V_{A}^{-1}\epsilon ^{1/3}k_{\perp }^{2/3}\mathcal{R}%
^{-1/6}\mathcal{L}^{2/3}\mathcal{L}^{\prime 1/6},  \label{A4} \\
\omega &=&C_{1}^{1/2}C_{2}\epsilon ^{1/3}k_{\perp }^{2/3}\mathcal{R}^{1/3}%
\mathcal{L}^{2/3}\mathcal{L}^{\prime -1/3},  \label{A5} \\
P_{\delta b_{\perp }} &=&C_{1}\epsilon ^{2/3}k_{\perp }^{-5/3}\mathcal{R}%
^{-1/3}\mathcal{L}^{-2/3}\mathcal{L}^{\prime 1/3},  \label{A6} \\
P_{\delta e_{\perp }} &=&C_{1}\epsilon ^{2/3}k_{\perp }^{-5/3}\mathcal{R}%
^{-4/3}\mathcal{L}^{-2/3}\mathcal{L}^{\prime -2/3}\left[ \left( 1+\rho
_{i}^{2}k_{\perp }^{2}\right) \mathcal{L}-\rho _{s}^{2}k_{\perp }^{2}\lambda
_{i}^{2}k_{z}^{2}\right] ^{2}.  \label{A7}
\end{eqnarray}

\section{Intermittent Turbulence}

Alfv\'{e}nic turbulence is in most cases intermittent (non space-filling),
and the turbulent fluctuations occupy only a fraction of volume (see papers
by Boldyrev \& Perez 2012 and TenBarge \& Howes 2013). The probability of
intermittent structures in the fluid turbulence is $p\left( l\right) \propto
l^{3-D}$, where the fractal dimensions are $D=0$, 1 and 2 for the ball-like,
tube-like and sheet-like structures (Frisch 1995). As the Alfv\'{e}nic
turbulent fluctuations are elongated in the direction of mean magnetic
field, the isotropic ball-like fluctuations can hardly be developed. For the
remaining two structure types we define the probability in the wavenumber
space instead of the space of scales: $p\left( k_{\perp }\right)
=C_{p}k_{\perp }^{-\alpha }$, where $\alpha =1$ and $2$ for the sheet-like
and tube-like structures, respectively. Then, using the energy flux in the
intermittent turbulence $\tilde{\epsilon}=p\left( k_{\perp }\right)
C_{1}^{-3/2}k_{\perp }\delta \tilde{v}_{e\perp }\delta \tilde{b}_{\perp
}^{2} $, the corresponding wave variables and spectral scalings are found:
\begin{eqnarray}
\delta \tilde{b}_{\perp } &=&C_{p}^{-1/3}k_{\perp }^{\alpha /3}\delta
b_{\perp },  \label{B1} \\
\delta \tilde{e}_{\perp } &=&C_{p}^{-1/3}k_{\perp }^{\alpha /3}\delta
e_{\perp },  \label{B2} \\
\delta \tilde{e}_{z} &=&C_{p}^{-2/3}k_{\perp }^{2\alpha /3}\delta e_{z},
\label{B3} \\
\tilde{k}_{z} &=&C_{p}^{-1/3}k_{\perp }^{\alpha /3}k_{z},  \label{B4} \\
\tilde{\omega} &=&C_{p}^{-1/3}k_{\perp }^{\alpha /3}\omega ,  \label{B5} \\
\tilde{P}_{\delta b_{\perp }} &=&C_{p}^{1/3}k_{\perp }^{-\alpha /3}P_{\delta
b_{\perp }},  \label{B6} \\
\tilde{P}_{\delta e_{\perp }} &=&C_{p}^{1/3}k_{\perp }^{-\alpha /3}P_{\delta
e_{\perp }}.  \label{B7}
\end{eqnarray}

This work was supported by the Belgian Federal Science Policy Office via
Solar-Terrestrial Centre of Excellence (project Fundamental Science) and via
IAP Programme (project P7/08 CHARM); by the European Commission via FP7
Program (project 313038 STORM), the NNSFC (11303099, 11373070, 11374262, and
41074107); by the MoSTC (grant 2011CB811402), the NSF of Jiangsu Province
(BK2012495), the Key Laboratory of Solar Activity at CAS NAO (LSA201304),
the CAEP, and the ITER-CN (2013GB104004).

\end{document}